\documentclass[12pt]{article}
\usepackage[]{amsmath,amssymb}
\numberwithin{equation}{section}
\newcounter{nt}
\newcommand{\note}{\refstepcounter{nt}\(^{\mbox{\footnotesize\arabic{nt}}}\)}
\DeclareMathOperator{\Imm}{Im}
\DeclareMathOperator{\supp}{supp}
\begin{document}
\title{NAIVE REALISM ABOUT OPERATORS}
\author{Martin Daumer, Detlef  D\"{u}rr\\ 
{\small Mathematisches Institut der Universit\"{a}t M\"{u}nchen}\\
{\small Theresienstra{\ss}e 39, 80333 M\"{u}nchen, Germany}
\and
        Sheldon  Goldstein\\ 
{\small Department of Mathematics, Rutgers University}\\
{\small New Brunswick, NJ 08903, USA}
\and
        Nino Zangh\`{\i}\\  
{\small Istituto di Fisica dell'Universit\`a di Genova, INFN} \\ 
{\small via Dodecaneso 33, 16146 Genova, Italy} } 
\date{January 11, 1996} 
\maketitle 

\begin{abstract} A source of much difficulty and confusion in the
interpretation of quantum mechanics is a ``naive realism about operators.''
By this we refer to various ways of taking too seriously the notion of
operator-as-observable, and in particular to the all too casual talk   about
``measuring operators'' that occurs when the subject is quantum mechanics.
Without a specification of what should be meant by ``measuring'' a quantum
observable, such an expression can have no clear meaning.  A definite
specification is provided by Bohmian mechanics, a theory that emerges from
Sch\"rodinger's equation for a system of particles when we merely insist
that ``particles'' means particles.  Bohmian mechanics clarifies the status
and the role of operators as observables in quantum mechanics by providing
the operational details absent from standard quantum mechanics. It thereby
allows us to readily dismiss all the radical claims traditionally enveloping
the transition from the classical to the quantum realm---for example, that
we must abandon classical logic or classical probability. The moral is
rather simple: Beware naive realism, especially about operators! 
\end{abstract} 

\section{Introduction} Traditional naive realism is the view that the world
{\it is\/} pretty  much the way it {\it seems,\/} populated by objects which
force  themselves upon our attention as, and which in fact are, the locus of
 sensual qualities.  A naive realist regards these ``secondary  qualities,''
for example color, as objective, as out there in the  world, much as
perceived.  A decisive difficulty with this view is  that once we
understand, say, how our perception of what we call color  arises, in terms
of the interaction of light with matter, and the  processing of the light by
the eye, and so on,\note\ we realize that  the presence out there of color
per se would play no role whatsoever  in these processes, that is, in our
understanding what is relevant to  our perception of ``color.'' At the same
time, we may also come to  realize that there is, in the description of an
object provided by the  scientific world-view, as represented say by
classical physics,  nothing which is genuinely ``color-like.''

We shall argue that the basic problem with quantum theory, more 
fundamental than the measurement problem and all the rest, is a naive 
realism about operators, a fallacy which we believe is far more 
serious than traditional naive realism: With the latter we are deluded 
partly by language but in the main by our senses, in a manner which 
can scarcely be avoided without a good deal of scientific or 
philosophical sophistication; with the former we are seduced by 
language alone, to accept a view which can scarcely be taken seriously 
without a large measure of (what often passes for) sophistication.

The classical physical observables---for a system of particles, their 
positions \(q=(\mathbf{q}_k)\), their momenta \(p=(\mathbf{p}_k)\), and 
the functions thereof, i.e., functions on phase space---form a 
commutative algebra.  It is generally taken to be the essence of 
quantization, the procedure which converts a classical theory to a 
quantum one, that \(q\), \(p\), and hence all  functions \(f(q,p)\) 
thereof are replaced by appropriate operators, on a Hilbert space, of 
possible wave functions, associated with the system under 
consideration.  Thus quantization leads to a noncommutative operator 
algebra of ``observables.'' Moreover, the fact that the observables in 
quantum theory form a noncommutative structure has traditionally been 
regarded as endowed with deep epistemological and or metaphysical 
significance and has variously been interpreted as the mathematical 
embodiment of irreducible indeterminacy or uncertainty and intrinsic 
fuzziness.

By naive realism about operators we refer to various, not entirely 
sharply defined, ways of taking too seriously the notion of 
operator-as-observable, and in particular to the all too casual talk 
about ``measuring operators'' which tends to occur as soon as a 
physicist enters quantum mode.  What, after all, is meant by measuring 
an operator?  If this is to have a meaning, that meaning must be 
supplied---it is not at all expressed by these words as they are 
normally understood.  But more on this later.

Not many physicists---or for that matter philosophers---have focused 
on the issue of naive realism about operators, but Schr\"odinger and 
Bell have expressed similar or related concerns:

\begin{quotation} \noindent\dots the new theory [quantum theory] \dots
considers the [classical] model  suitable for guiding us as to just which
measurements can in principle  be made on the relevant natural object. \dots
Would it not be pre-established harmony of a peculiar sort if the
classical-epoch  researchers, those who, as we hear today, had no idea of
what {\it  measuring\/} truly is, had unwittingly gone on to give us as
legacy a  guidance scheme revealing just what is fundamentally measurable
for  instance about a hydrogen atom!?  (Schr\"odinger 1935)
\end{quotation}

\begin{quotation}
Here are some words which, however legitimate and necessary in 
application, have no place in a {\it formulation\/} with any 
pretension to physical precision: {\it system; apparatus; environment; 
microscopic, macroscopic; reversible, irreversible; observable; 
information; measurement.\/}

 \dots The notions of ``microscopic'' and ``macroscopic'' defy precise 
definition. \dots  Einstein said that it is theory which decides what is 
``observable''.  I think he was right. \dots   ``observation'' is a 
complicated and theory-laden business.  Then that notion should not 
appear in the {\it formulation\/} of fundamental theory. \dots 

On this list of bad words from good books, the worst of all is 
``measurement''.  It must have a section to itself. (Bell 1990)
\end{quotation}

We agree almost entirely with Bell here.  We insist, however, that 
``observable'' is just as bad as ``measurement,'' maybe even a little 
worse.  Be that as it may, after listing Dirac's measurement 
postulates Bell continues:

\begin{quotation}
It would seem that the theory is exclusively concerned about ``results 
of measurement'', and has nothing to say about anything else.  What 
exactly qualifies some physical systems to play the role of 
``measurer''?  Was the wavefunction of the world waiting to jump for 
thousands of millions of years until a single-celled living creature 
appeared?  Or did it have to wait a little longer, for some better 
qualified system \dots   with a Ph.D.?  If the theory is to apply to 
anything but highly idealized laboratory operations, are we not 
obliged to admit that more or less ``measurement-like'' processes are 
going on more or less all the time, more or less everywhere.  Do we 
not have jumping then all the time?

The first charge against ``measurement'', in the fundamental axioms of 
quantum mechanics, is that it anchors the shifty split of the world 
into ``system'' and ``apparatus''.  A second charge is that the word 
comes loaded with meaning from everyday life, meaning which is 
entirely inappropriate in the quantum context.  When it is said that 
something is ``measured'' it is difficult not to think of the result 
as referring to some {\it preexisting property\/} of the object in 
question.  This is to disregard Bohr's insistence that in 
quantum phenomena the apparatus as well as the system is essentially 
involved.  If it were not so, how could we understand, for example, 
that ``measurement'' of a component of ``angular momentum''  \dots {\it in 
an arbitrarily chosen direction\/}  \dots   yields one of a discrete set 
of values?  When one forgets the role of the apparatus, as the word 
``measurement'' makes all too likely, one despairs of ordinary 
logic \dots hence ``quantum logic''.  When one remembers the role of the 
apparatus, ordinary logic is just fine.

In other contexts, physicists have been able to take words from 
ordinary language and use them as technical terms with no great harm 
done.  Take for example the ``strangeness'', ``charm'', and ``beauty'' 
of elementary particle physics.  No one is taken in by this ``baby 
talk''. \dots   Would that it were so with ``measurement''.  But in fact 
the word has had such a damaging effect on the discussion, that I 
think it should now be banned altogether in quantum mechanics.
({\sl Ibid.\/})
\end{quotation}

While Bell focuses directly here on the misuse of the word  ``measurement''
rather than on that of ``observable,'' it is worth noting that the abuse of
``measurement'' is in a sense inseparable  from that of ``observable,''
i.e., from naive realism about operators.   After all one would not be very
likely to speak of measurement unless  one thought that something, some
``observable'' that is, was somehow  there to be measured.

More Bell:

\begin{quotation}
The concept of `measurement' becomes so fuzzy on reflection that it is 
quite surprising to have it appearing in physical theory {\it at the 
most fundamental level.\/} Less surprising perhaps is that 
mathematicians, who need only simple axioms about otherwise undefined 
objects, have been able to write extensive works on quantum 
measurement theory---which experimental physicists do not find it 
necessary to read. \dots   Does not any {\it analysis\/} of measurement 
require concepts more {\it fundamental\/} than measurement?  And 
should not the fundamental theory be about these more fundamental 
concepts?  (Bell 1981)
\end{quotation}

\begin{quotation}
\noindent \dots in physics the only observations we must consider are position 
observations, if only the positions of instrument pointers.  It is a 
great merit of the de Broglie-Bohm picture to force us to consider 
this fact.  If you make axioms, rather than definitions and theorems, 
about the `measurement' of anything else, then you commit redundancy 
and risk inconsistency.  (Bell 1982)
\end{quotation}

If our feeling that Bell's words are thoroughly compelling were widely 
shared, by physicists and philosophers, there would perhaps be little 
point in continuing this paper.  But it is not, so we continue!  
Moreover, we wish in any case to focus on what Bell calls ``the 
de~Broglie-Bohm picture''---what we prefer to call Bohmian 
mechanics---for the light it sheds on naive realism about operators.

We wish to do two things here: We wish to elaborate on why we think 
what we have called naive realism about operators (taking operators 
too seriously as observables) is bad, and we wish to relate this issue 
to Bohmian mechanics.  Briefly stated, the relevant connections 
between naive realism about operators and Bohmian mechanics are the following:
\begin{enumerate}

     \item A frequent complaint about Bohmian mechanics, in which positions 
play a fundamental role, is expressed in terms of questions like 
``What about other observables?''

     \item Bohmian mechanics allows the inadequacy, indeed the utter
wrongheadedness, of naive realism about operators to emerge with stark
clarity.

\end{enumerate}

\section{Bohmian mechanics}

According to orthodox quantum theory, the {\it complete\/} description 
of a system of particles is provided by its wave function.  It is 
rarely noticed that even this statement is somewhat problematical: If 
``particles'' is intended with its usual meaning---point-like entities 
whose most important feature is their positions in space---the 
statement is clearly false, since the complete description would then 
have to include these positions; otherwise, the statement is, to be 
charitable, vague.  Bohmian mechanics is the theory which emerges when 
we indeed insist that ``particles'' means particles.

According to Bohmian mechanics (Bohm 1952, Bohm et al.  1993, Bell 
1987, D\"urr et al.  1992, 1996, Holland 1993, Berndl et al. 1995), 
the complete description of an \(N\)-particle system is provided by 
its wave function \(\psi\) {\it and\/} its configuration 
\(Q=(\mathbf{Q}_{1},\dots,\mathbf{Q}_{N}),\) where the 
\(\mathbf{Q}_k\) are the positions of the particles.  The wave 
function, which evolves according to Schr\"odinger's equation, 
choreographs the motion of the particles: these evolve---in the 
simplest manner possible---according to a first-order ordinary 
differential equation

\begin{equation*}
	\frac{dQ}{dt}=v^{\psi}(Q)
\end{equation*}
whose right-hand side, a velocity vector field on configuration space, 
is generated by the wave function.  Considerations of space-time 
symmetry---Galilean and time-reversal invariance---then determine the 
form of 
\(v^{\psi}= (\mathbf{v}^{\psi}_{1},\dots,\mathbf{v}^{\psi}_{N})\) 
(D\"urr et al.  1992), and we arrive at the defining (evolution) 
equations of {\it Bohmian mechanics\/}:
\begin{align}
	\frac{d\mathbf{Q}_{k}}{dt}&= 
	\mathbf{v}^{\psi}_{k}(\mathbf{Q}_{1},\dots,\mathbf{Q}_{N})
	\equiv\frac{\hbar}{m_k}
	\Imm\frac{\boldsymbol{\nabla}_{\mathbf{q}_{k}}\psi}{\psi}
	(\mathbf{Q}_{1},\dots,\mathbf{Q}_{N})
	\label{guide}\\
	\intertext{and}
	i\hbar\frac{\partial\psi}{\partial t}&= H\psi
	\label{se}
\end{align}
where \(H\) is the usual Schr\"odinger Hamiltonian, containing as parameters 
the masses \(m_{1},\dots,m_{N}\) of the particles as well as the potential 
energy function \(V\) of the system.

For an \(N\)-particle universe, these two equations form a complete 
specification of the theory.  There is no need, and indeed no room, 
for any further axioms, describing either the behavior of ``other 
observables'' or the effects of ``measurement.''
\medskip

Bohmian mechanics is the most naively obvious embedding imaginable of 
Schr\"odinger's equation into a completely coherent physical theory!  
If one didn't already know better, one would naturally conclude that 
it can't ``work,'' i.e., that it can't account for quantum phenomena.  
After all, if something so obvious and, indeed, so trivial works, 
great physicists would never have insisted, as they have and as they 
continue to do, that quantum theory demands radical epistemological 
and metaphysical innovations.

Moreover, when we think about it, how {\it could\/} Bohmian mechanics 
have much to do with quantum theory?  Where is quantum randomness in 
this deterministic theory?  Where is quantum uncertainty?  Where are 
operators as observables and all the rest?

Be that as it may, Bohmian mechanics is certainly {\it a\/} theory.   It
describes a world in which particles move in a highly non-Newtonian  sort of
way, and it would do so even were it the case that the way they do move in
this theory had absolutely nothing to do with quantum mechanics.

It turns out, however, that a remarkable consequence (!) of the 
equations \eqref{guide} and \eqref{se} is that when a system has wave 
function \(\psi\) its configuration is random, with probability 
density \(\rho\) given typically by \(\rho=|\psi|^{2}\), the {\it quantum 
equilibrium\/} distribution.  In other words, it turns out that 
systems are somehow typically {\it in quantum equilibrium.\/} Moreover, this 
conclusion comes together with the clarification of what precisely 
this means, and also implies that a Bohmian universe embodies an 
absolute uncertainty which can itself be regarded as the origin of the 
uncertainty principle.  We shall not go into these matters here, 
having discussed them at length elsewhere (D\"urr et al. 1992, 1996). We 
note, however, that nowadays, with chaos theory and nonlinear dynamics 
so fashionable, it is not generally regarded as terribly astonishing 
for an appearance of randomness to emerge from a deterministic 
dynamical system.

It also turns out that the entire quantum formalism, operators as 
observables and all the rest, is a consequence of Bohmian mechanics, 
and since this is relevant to the issue of naive realism about 
operators, we do wish to spend some time sketching how this comes 
about.

\section{The quantum formalism}

Information about a system does not spontaneously pop into our heads, or
into our (other) ``measuring'' instruments; rather, it is generated by an
{\it experiment:\/} some physical interaction between the system of
interest and these instruments, which together (if there is more than one)
comprise the {\it apparatus\/} for the experiment.  Moreover, this
interaction is defined by, and must be analyzed in terms of, the physical
theory governing the behavior of the composite formed by system and
apparatus. If the apparatus is well designed, the experiment should somehow
convey significant information about the system. However, we cannot hope to
understand the significance of this ``information''---for example, the
nature of what it is, if anything, that has been measured---without some
such theoretical analysis.

Whatever its significance, the information conveyed by the experiment is
registered in the apparatus as an {\it output\/}, represented, say, by the
orientation of a pointer. Moreover, when we speak of an experiment, we have
in mind a fairly definite initial state of the apparatus, the ready state,
one for which the apparatus should function as intended, and in particular
one in which the pointer has some ``null'' orientation.

For Bohmian mechanics we should expect in general that, as a  consequence of
the quantum equilibrium hypothesis, the outcome of the 
ex\-per\-i\-ment---the final point\-er ori\-en\-tation---will be random:
Even if  the system-apparatus composite initially has a definite, known wave
function,  so that the outcome is determined by the initial configuration of
 system and apparatus, this configuration is random, since the  composite
system is in quantum equilibrium, i.e., the distribution of  this
configuration is given by \(|\Psi(x,y)|^2\), where \(\Psi\) is the  wave
function of the system-apparatus composite and \(x\), respectively  \(y\), is
the generic system, respectively apparatus, configuration.  There  are,
however, special experiments whose outcomes are somewhat less  random than
we might have thought possible.

In fact, consider a {\it measurement-like\/} experiment, one which is {\it
reproducible\/} in the sense that it will yield the same outcome as
originally obtained if it is immediately repeated. (Note that this means
that the apparatus must be immediately reset to its ready state, or a
fresh apparatus must be employed, while the system is not tampered with so
that its initial state for the repeated experiment is its final state
produced by the first experiment.) Suppose that this experiment admits,
i.e., that the apparatus is so designed that there are, only a finite (or
countable) number of possible outcomes \(\alpha\),\note\ for example,
\(\alpha=\)``left'' and \(\alpha=\)``right''.
The experiment also usually comes equipped with a {\it calibration\/}
\(\lambda_\alpha\), an assignment of numerical values (or a vector of such
values) to the various outcomes \(\alpha\).

It can be shown (Daumer et al. 1996), under 
further simplifying assumptions, that for such reproducible 
experiments there are special subspaces \(\mathcal{H}_\alpha\) of the 
system Hilbert space \(\mathcal{H}\) of (initial) wave functions, 
which are mutually orthogonal and span the entire system Hilbert space

\begin{equation}
\mathcal{H}=\bigoplus_\alpha{\mathcal{H}_\alpha},	
	\label{Hdecomp}
\end{equation}
such that if the system's wave function is initially in 
\(\mathcal{H}_\alpha\), outcome \(\alpha\) definitely occurs and the value 
\(\lambda_\alpha\) is thus definitely obtained.  It then follows that 
for a general initial system wave function 
\begin{equation}
\psi=\sum_\alpha\psi_\alpha\equiv\sum_\alpha{P_{\mathcal{H}_\alpha}}\psi
    \label{psidecomp}
\end{equation}
where \(P_{\mathcal{H}_\alpha}\) is the projection onto the subspace 
\(\mathcal{H}_\alpha\), the outcome \(\alpha\) is obtained with (the 
usual) probability\note\
\begin{equation}
     p_\alpha={\Vert P_{\mathcal{H}_\alpha}\psi\Vert}^2.
    \label{prob}
\end{equation}     
In particular, the expected value obtained is
\begin{equation}
    \sum_\alpha{p_\alpha \lambda_\alpha}=\sum_\alpha{\lambda_\alpha{\Vert
     P_{\mathcal{H}_\alpha}\psi\Vert}^2}=\langle\psi,A\psi\rangle
   \label{ev}
\end{equation} 
where
\begin{equation}
     A=\sum_{\alpha}{\lambda_\alpha P_{\mathcal{H}_\alpha}}
  \label{A}
\end{equation} 
and \(\langle\,\cdot\,,\,\cdot\,\rangle\) is the usual inner product:
\begin{equation}
     \langle\psi,\phi\rangle=\int{{\psi}^*(x)\,\phi(x)\,dx}.
  \label{ip}
\end{equation} 

What we wish to emphasize here is that, insofar as the statistics for 
the values which result from the experiment are concerned, the 
relevant data for the experiment are the collection 
(\(\mathcal{H}_\alpha\)) of special subspaces, together with the 
corresponding calibration (\(\lambda_\alpha\)), and {\it this data is 
compactly expressed and represented by the self-adjoint operator 
\(A\), on the system Hilbert space \(\mathcal{H}\), given by 
\eqref{A}.\/} Thus with a reproducible experiment \(\mathfrak{E}\) we 
naturally associate an operator \(A=A_{\mathfrak{E}}\), %

\begin{equation}
     \mathfrak{E}\mapsto A,
   \label{EA}
\end{equation} 
a single mathematical object, defined on the system alone, in terms of
which an efficient description of the possible results is achieved. If we
wish we may speak of operators as observables, but if we do so it is
important that we appreciate that in so speaking we merely refer to what we
have just sketched: the role of operators in the description of certain
experiments.\note\

In particular, so understood, the notion of operator-as-observable in no
way implies that anything is measured in the experiment, and certainly not
the operator itself!  In a general experiment no system property is being
measured, even if the experiment happens to be measurement-like. (Position
measurements are of course an important exception.) What in general is
going on in obtaining outcome \(\alpha\) is completely straightforward and
in no way suggests, or assigns any substantive meaning to, statements to
the effect that, prior to the experiment, observable \(A\) somehow had a
value \(\lambda_\alpha\)---whether this be in some determinate sense or in
the sense of Heisenberg's ``potentiality'' or some other ill-defined fuzzy
sense---which is revealed, or crystallized, by the experiment.\note\

Much of the preceding sketch of the emergence and role of operators as 
observables in Bohmian mechanics, including of course the von 
Neumann-type picture of ``measurement'' at which we arrive, applies as 
well to orthodox quantum theory. In fact, it would appear that the 
argument against naive realism about operators provided by such an 
analysis has even greater force from an orthodox perspective: Given 
the initial wave function, at least in Bohmian mechanics the outcome 
of the particular experiment is determined by the initial 
configuration of system and apparatus, while for orthodox quantum 
theory there is nothing in the initial state which completely 
determines the outcome.  Indeed, we find it rather surprising that 
most proponents of the von Neumann analysis of measurement, beginning 
with von Neumann, nonetheless seem to retain their naive realism about 
operators.  Of course, this is presumably because more urgent 
matters---the measurement problem and the suggestion of inconsistency 
and incoherence that it entails---soon force themselves upon one's 
attention.  Moreover such difficulties perhaps make it difficult to 
maintain much confidence about just what {\it should\/} be concluded 
{}from the ``measurement'' analysis, while in Bohmian mechanics, for 
which no such difficulties arise, what should be concluded is rather 
obvious.\note\

\section{The reality of spin and other observables}

The canonical example of a ``quantum measurement'' is provided by the
Stern-Gerlach experiment. We wish to focus on this example here in order to
make our previous considerations more concrete, as well as to present some
further considerations about the ``reality'' of operators-as-observables.
We wish in particular to comment on the status of spin. We shall therefore
consider a Stern-Gerlach ``measurement'' of the spin of an electron, even
though such an experiment is generally believed to be unphysical (Mott
1929), rather than of the internal angular momentum of a neutral atom.

We must first explain how to incorporate spin into Bohmian mechanics.  
This is very easy; we need do, in fact, almost nothing: Our derivation 
of Bohmian mechanics (D\"urr et al. 1992) was based in part on 
rotation invariance, which requires in particular that rotations act 
on the value space of the wave function.  The latter is rather 
inconspicuous for spinless particles---with complex-valued wave 
functions, what we have been considering up till now---since rotations 
then act in a trivial manner on the value space \(\mathbb{C}\).  The 
simplest nontrivial (projective) representation of the rotation group 
is the \(2\)-dimensional, ``spin \(\frac{1}{2}\)'' representation; 
this representation leads to a Bohmian mechanics involving 
spinor-valued wave functions for a single particle and 
spinor-tensor-product-valued wave function for many particles.  Thus 
the wave function of a single spin--\(\frac{1}{2}\) particle has two 
components 
\begin{equation}
	  \psi(\mathbf{q}) = \begin{pmatrix} \psi_{1}(\mathbf{q}) \\ 
	  \psi_{2}(\mathbf{q})\end{pmatrix}
	\label{sp}
\end{equation}
which get mixed under rotations according to the action generated by the
Pauli spin matrices \(\boldsymbol{\sigma}=(\sigma_x,\sigma_y,\sigma_z)\), which
may be taken to be
\begin{equation}
	\sigma_x= \begin{pmatrix}0&1\\1&0\end{pmatrix}
	 \quad
	 \sigma_y= \begin{pmatrix}0&-i\\i&0\end{pmatrix}
	 \quad
	 \sigma_z= \begin{pmatrix}1&0\\0&-1\end{pmatrix}
	\label{pm}
\end{equation}
Beyond the fact that the wave function now has a more abstract value 
space, nothing changes from our previous description: The wave 
function evolves via \eqref{se}, where now the Hamiltonian \(H\) 
contains the Pauli term, for a single particle proportional to 
\(\mathbf{B}\cdot\boldsymbol{\sigma}\), which represents the coupling 
between the ``spin'' and an external magnetic field \(\mathbf{B}\).  
The configuration evolves according to the natural extension of 
\eqref{guide} to spinors, obtained say by multiplying both the 
numerator and denominator of the argument of ``\(\Imm\)'' on the left by 
\(\psi^*\) and interpreting the result for the case of spinor values 
as a spinor-inner-product.

Let's focus now on a Stern-Gerlach ``measurement of \(A=\sigma_z\).'' 
An inhomogeneous magnetic field is established in a neighborhood of 
the origin, by means of a suitable arrangement of magnets.  This 
magnetic field is oriented more or less in the positive 
\(z\)-direction, and is increasing in this direction.  We also assume 
that the arrangement is invariant under translations in the 
\(x\)-direction, i.e., that the geometry does not depend upon 
\(x\)-coordinate.  An electron, with a fairly definite momentum, is 
directed towards the origin along the negative \(y\)-axis.  Its 
passage through the inhomogeneous field generates a vertical 
deflection of its wave function away from the \(y\)-axis, which for 
Bohmian mechanics leads to a similar deflection of the electron's 
trajectory.  If its wave function \(\psi\) were initially an 
eigenstate of \(\sigma_z\) of eigenvalue \(1\) (\(-1\)), i.e., if it 
were of the form
\begin{equation}
	\psi=|\uparrow\,\rangle\otimes\phi_0
	\quad(\psi=|\downarrow\,\rangle\otimes\phi_0)
	\label{updown}
\end{equation}
where

\begin{equation}
	|\uparrow\,\rangle= \begin{pmatrix}1\\0\end{pmatrix}
	\quad\text{and}\quad
	|\downarrow\,\rangle= \begin{pmatrix}0\\1\end{pmatrix},
	\label{sb}
\end{equation}
then the deflection would be in the positive (negative) 
\(z\)-direction (by a rather definite angle).  For a more general 
initial wave function, passage through the magnetic field will, by 
linearity, split the wave function into an upward-deflected piece 
(proportional to \(|\uparrow\,\rangle\)) and a downward-deflected 
piece (proportional to \(|\downarrow\,\rangle\)), with corresponding 
deflections of the possible trajectories.

The outcome is registered by detectors placed in the way of these two
``beams.'' Thus of the four kinematically possible outcomes (``pointer
positions'') the occurrence of no detection defines the null output,
simultaneous detection is irrelevant ( since it does not occur if the
experiment is performed one particle at a time), and the two relevant
outcomes correspond  to registration by either the upper or the
lower detector. Thus the calibration for a measurement of \(\sigma_z\) is
\(\lambda_{\text{up}}=1\) and \(\lambda_{\text{down}}=-1\) (while for a
measurement of the \(z\)-component of the spin angular momentum itself the
calibration is the product of what we have just described by 
\(\frac{1}{2}\hbar\)).

Note that one can completely understand what's going on in this 
Stern-Gerlach experiment without invoking any additional property of 
the electron, e.g., its {\it actual\/} \(z\)-component of spin that is 
revealed in the experiment.  For a general initial wave function there 
is no such property; what is more, the transparency of the analysis of 
this experiment makes it clear that there is nothing the least bit 
remarkable (or for that matter ``nonclassical'') about the {\it 
nonexistence\/} of this property.  As we emphasized earlier, it is 
naive realism about operators, and the consequent failure to pay 
attention to the role of operators as observables, i.e., to precisely 
what we should mean when we speak of measuring operator-observables, 
that creates an impression of quantum peculiarity.

Bell has said that (for Bohmian mechanics) spin is not real.  Perhaps 
he should better have said: {\it ``Even\/} spin is not real,'' not 
merely because of all observables, it is spin which is generally 
regarded as quantum mechanically most paradigmatic, but also because 
spin is treated in orthodox quantum theory very much like position, as 
a ``degree of freedom''---a discrete index that supplements the 
continuous degrees of freedom corresponding to position---in the wave 
function.  Be that as it may, his basic meaning is, we believe, this: 
Unlike position, spin is not {\it primitive\/},\note\ i.e., no {\it 
actual\/} discrete degrees of freedom, analogous to the {\it actual\/} 
positions of the particles, are added to the state description in 
order to deal with ``particles with spin.'' Roughly speaking, spin is 
{\it merely\/} in the wave function.  At the same time, as just said, 
``spin measurements'' are completely clear, and merely reflect the way 
spinor wave functions are incorporated into a description of the 
motion of configurations.

It might be objected that while spin may not be primitive, so that the result
of our ``spin measurement'' will not reflect any initial primitive property
of the system, nonetheless this result {\it is\/} determined by the initial
configuration of the system, i.e., by the position of our electron,
together with its initial wave function, and as such---as a function
\(X_{\sigma_z}(\mathbf{q}, \psi)\) of the state of
the system---it is  some property of the system and in particular it is
surely real. Concerning this, several comments.

The function \(X_{\sigma_z}(\mathbf{q}, \psi)\), or better the 
property it represents, is (except for rather special choices of 
\(\psi\)) an extremely complicated function of its arguments; it is 
not ``natural,'' not a ``natural kind'': It is not something in which, 
in its own right, we should be at all interested, apart from its 
relationship to the {\it result\/} of this particular experiment.

Be that as it may, it is not even possible to identify this function 
\(X_{\sigma_z}(\mathbf{q}, \psi)\) with the measured spin component, 
since different experimental setups for ``measuring the spin 
component'' may lead to entirely different functions.  In other words 
\(X_{\sigma_z}(\mathbf{q}, \psi)\) is an abuse of notation, since the 
function \(X\) should be labeled, not by \(\sigma_z\), but by the 
particular experiment for ``measuring \(\sigma_z\)''.

For example (Albert 1992, p.153), if \(\psi\) and the magnetic field 
have sufficient reflection symmetry with respect to a plane between 
the poles of our SG magnet, and if the magnetic field is reversed, 
then the sign of what we have called \(X_{\sigma_z}(\mathbf{q}, 
\psi)\) will be reversed: for both orientations of the magnetic field 
the electron cannot cross the plane of symmetry and hence if initially 
above respectively below the symmetry plane it remains above 
respectively below it.  But when the field is reversed so must be the 
calibration, and what we have denoted by \(X_{\sigma_z}(\mathbf{q}, 
\psi)\) changes sign with this change in experiment.\note\  

In general \(X_A\) does not exist, i.e., \(X_{\mathfrak{E}}\), the 
result of the experiment \(\mathfrak{E}\), in general depends upon 
\(\mathfrak{E}\) and not just upon \(A=A_{\mathfrak{E}}\), the 
operator associated with \(\mathfrak{E}\).  In foundations of quantum 
mechanics circles this situation is referred to as {\it 
contextuality,\/} but we believe that this terminology, while quite 
appropriate, somehow fails to convey with sufficient force the rather 
definitive character of what it entails: Properties that are merely 
contextual are not properties at all; they do not exist, and their 
failure to do so is in the strongest sense possible!  We thus believe 
that contextuality reflects little more than the rather obvious 
observation that the result of an experiment should depend upon how it 
is performed!

We summarize our comparison of the status of position with that of 
other observables in the following chart:
{\large  
$$\begin{tabular}{|c|c|}
	\hline
	position & other observables  \\
	\hline\hline
	real & not real  \\
	\hline
	primitive & not primitive  \\
	\hline
	natural (kind) & not natural (kind)  \\
	\hline
	noncontextual & contextual  \\
	\hline \end{tabular}$$ } 

\section{Hidden variables} What about the ``no-go'' theorems for hidden
variables?\note\ These  theorems show that there is no ``good'' map
$A\mapsto X_{A}$ from  operators to random variables (on the space of
``hidden variables''), where by ``good'' we mean in the sense  that the joint
distributions of the random variables are consistent  with the corresponding
quantum mechanical distributions whenever the  latter are defined.  

As commonly understood, these theorems involve a certain irony: They 
conclude with the impossibility of a deterministic description, or  more
generally of any sort of realist description, only by in effect  themselves
assuming a ``realism'' of a most implausible variety,  namely, naive realism
about operators. For why else would a realist, even one who is also a
determinist, expect there to be such a map? After all, the fact that the
same operator plays a role in different experiments  does not imply that
these experiments have much else in common, and  certainly not that they
involve measurements of the same  thing. It is thus with detailed
experiments, and not with the associated operators, that random variables
might reasonably be expected to be associated.

When faced with the inconsistency of possible values as expressed by the
``no-go'' theorems, how should one  respond?  As does a ``typical'' physicist,
by declaring in effect that  quantum mechanics does not allow us to ask the
obvious questions?  But  even if we should chose to forbid ourselves from
asking sufficiently many questions to notice  it, the state of affairs
described by the theorems nontheless  logically implies the obvious
conclusion, namely, that the  incompatible joint values refer to different,
and incompatible,  experimental set-ups, just as Bohr  told us all along.
This  mathematical incompatibility of ``joint values'' thus seems to attain 
genuine physical significance only to the extent that we are seduced  by
naive realism about operators.\note\

Referring to the axioms involved in the no-hidden-variables theorems, 
Bell says:
\begin{quote}
A final moral concerns terminology. Why did such serious
people take so
seriously axioms which now seem so arbitrary? (Bell 1982)	
\end{quote}
To this question we are tempted to respond that the answer, of course, 
is that these ``serious people'' were deluded by naive realism about 
operators.  However, what Bell is really asking is why they should 
have been so deluded, as is made clear by what he says next:
\begin{quote}
I suspect that they were misled by the pernicious misuse of the word 
`measurement' in contemporary theory.  This word very strongly 
suggests the ascertaining of some preexisting property of some thing, 
any instrument involved playing a purely passive role.  Quantum 
experiments are just not like that, as we learned especially from 
Bohr.  The results have to be regarded as the joint product of 
`system' and `apparatus,' the complete experimental set-up.  But the 
misuse of the word `measurement' makes it easy to forget this and then 
to expect that the `results of measurements' should obey some simple 
logic in which the apparatus is not mentioned.  The resulting 
difficulties soon show that any such logic is not ordinary logic.
\end{quote}
\noindent Note, in particular, the sentence that ends with ``in which 
the apparatus is not mentioned.'' This makes little sense without an 
implicit reference to naive realism about operators: Everyone would 
agree that, even if it were not necessary to mention the apparatus 
{\it per se\/}, at least {\it something\/} would have to be mentioned; 
Bell is here criticizing the view that for this ``something,'' the 
operator-as-observable that is being ``measured'' should suffice.

Bell continues:
\begin{quote}
It is my impression that the whole vast subject of `Quantum Logic' has 
arisen in this way from the misuse of a word.  I am convinced that the 
word `measurement' has now been so abused that the field would be 
significantly advanced by banning its use altogether, in favour for 
example of the word `experiment.'
\end{quote}

\section{Comments}

Let's reconsider a fragment of one of our previous Bell
quotations:
\begin{quote}
If you make axioms, rather than definitions and theorems, about the 
`measurement' of anything else [other than position], then you commit 
redundancy and risk inconsistency.
\end{quote}
\noindent We would like to propose what we believe to be a small 
improvement.  Replace ``measurement'' by ``behavior.'' Then add to 
``redundancy'' and ``inconsistency'' the further possibility of 
irrelevance.  In other words we are proposing the following amendment:
\begin{quote}
\textsl{If you make axioms, rather than definitions and theorems, 
about the behavior of anything else---beyond what is required to fully 
specify the behavior of positions---then either you commit redundancy 
and risk inconsistency, or you commit irrelevancy}
\end{quote}
For example, suppose we add, say to Bohmian mechanics, some axioms 
governing the behavior of ``momentum.'' Then there are two 
possibilities:
\begin{enumerate}
     
     \item On the one hand, by ``momentum'' we may mean, say, mass times 
      velocity, in which case we have either redundancy or inconsistency.  

     \item On the other hand, if ``momentum'' is not given a meaning in 
      terms of the behavior of configurations---if it is a brand new 
      property as it were---then it is irrelevant!

\end{enumerate}
\bigskip

A related lesson of Bohmian mechanics is one of flexibility: Not only  need
we not consider ``other observables'' on a fundamental level, it  is not
even necessary that the primitive variables (what the theory is 
fundamentally {\it about\/})---in Bohmian mechanics the positions of  the
particles---be ``observables'' in the sense that they are  associated with
self-adjoint operators in the usual way.  That they  are for Bohmian
mechanics is best regarded as an accident arising from  incidental features
(for example, the form of the inner product) of the mathematical structure
of nonrelativistic  quantum theory.\note\ \bigskip

\bigskip

Let's now return to the objection, ``What about other observables?'' 
Since operators as observables are nothing more than a convenient 
mathematical device for describing what is most relevant about certain 
special experiments, asking this question amounts to nothing more than 
asking, ``What about special experiments?'' But put this way, there is 
no longer any suggestion of inadequacy or incompleteness.

\section{Other interpretations}

In this brief section we wish to  outline how some of the more familiar 
interpretations of quantum theory fare with regard to the fallacy of 
naive realism about operators. We do so in the following chart:
{\large $$  
\begin{tabular}{|c|c|}
	\hline
	guilty &   not guilty\\
	\hline\hline
	Copenhagen (quantum orthodoxy) &  Copenhagen   (Bohr)\\
	\hline
	many worlds\note\  &   many worlds\note\ \\
	\hline
	quantum logic\note\ &   many minds\note\ \\
	\hline
	 quantum probability\note\ &  spontaneous localization\note\ \\
	\hline
	modal interpretation\note\ &  stochastic mechanics\note\ \\
	\hline
	consistent histories\note\  &  Bohmian mechanics\\
	\hline
\end{tabular}
$$} 
Note that, as is so often the case, the Copenhagen interpretation 
is hard to pin down!

\section{Diatribe}
Why should we (continue to) insist upon a metaphysics---that 
observables or properties should somehow be identified with 
operators---which, while seeming to express the essential innovation 
of quantum theory, in fact conflicts (or at least is strongly at odds 
with) the very mathematical structure of the theory itself?  What is 
the point of multiplying properties, new properties irreducible to 
what we have already, when their mutual incompatibility has been 
enshrined in quantum orthodoxy from its very inception (the 
uncertainty principle, complementarity); when the no-hidden-variables 
theorems establish their joint impossibility; so that in order to save 
them one must resort to such expedients, contortions, and perversions 
as quantum logic and quantum probability (or, at best, to something 
like van Fraassen's modal interpretation of quantum theory (van 
Fraassen 1991), with all the enormous complexity its formulation 
requires); when they add nothing of substance or value to our 
understanding of the use of operators as ``observables''---of the role 
of operators in quantum theory---which is in fact quite 
straightforward, as a compact expression of the most important or 
relevant features of certain experiments, the analysis of which 
reveals that what is going on during such experiments is in general 
not a measurement of the associated operator---what would that mean 
anyway?---nor, indeed, of anything else worth mentioning!?

\section*{Acknowledgments} We are grateful to Rebecca Goldstein for her
suggestions. This work was supported in part by NSF Grant No. DMS--9504556,
by the DFG, and by the INFN.
\bigskip

{\footnotesize 
\section*{Notes}
\setcounter{nt}{0}

\noindent\note\ That we don't understand the last link in the causal chain 
leading to our {\it conscious\/} perception is not very relevant here.
\medskip

\noindent\note\ This is really no assumption at all, since the outcome
should ultimately be converted to digital form, whatever its initial
representation may be.
\medskip

\noindent\note\ In the simplest such situation the unitary evolution 
for the wave function of the composite system carries the initial wave 
function \(\Psi_i=\psi\otimes\Phi_0\) to the final wave function 
\(\Psi_f=\sum_\alpha\psi_{\alpha}\otimes\Phi_\alpha\), where 
\(\Phi_0\) is the ready apparatus wave function, and \(\Phi_\alpha\) 
is the apparatus wave function corresponding to outcome \(\alpha\).  
Then integrating \({|\Psi_f|}^2\) over \(\supp\Phi_\alpha\), we 
immediately arrive at \eqref{prob}.  
\medskip

\noindent\note\ Operators as observables also naturally convey 
information about the system's wave function after the experiment.  
For example, for an ideal measurement, when the outcome is \(\alpha\) 
the wave function of the system after the experiment is (proportional 
to) \(P_{\mathcal{H}_{\alpha}}\psi\).  
\medskip

\noindent \note\ Even speaking of the observable \(A\) as having value
\(\lambda_\alpha\) when the system's wave function is in 
\(\mathcal{H}_{\alpha}\), i.e., when this wave function is an eigenstate 
of \(A\) of eigenvalue \(\lambda_\alpha\), insofar as it suggests that 
something peculiarly quantum is going on when the wave function is not 
an eigenstate whereas in fact there is nothing the least bit peculiar 
about the situation, perhaps does more harm than good.
\medskip

\noindent\note\ It might be objected that we are claiming to arrive at  the
quantum formalism under somewhat unrealistic assumptions, such as,  for
example, reproducibility.  (We note in this regard that many more 
experiments than those satisfying our assumptions can be associated  with
operators in exactly the manner we have described.) We agree.   But this
objection misses the point of the exercise.  The quantum  formalism itself
is an idealization; when applicable at all, it is  only as an approximation.
 Beyond illuminating the role of operators  as ingredients in this
formalism, our point was to indicate how  naturally it emerges.  In this
regard we must emphasize that the  following question arises for quantum
orthodoxy, but does not arise  for Bohmian mechanics: For precisely which
theory is the quantum  formalism an idealization?  (For further  elaboration
on this point, as  well as for a discussion of how ``generalized
observables'' (Davies,  1976) naturally arise in Bohmian mechanics, see 
D\"urr et al. 1996 and Daumer et al. 1996.) \medskip

\noindent\note\ We should probably distinguish two senses of
``primitive'': i) the {\it strongly primitive\/} variables, which describe
what the theory is fundamentally {\it about\/}, and ii) the {\it weakly
primitive\/} variables, the basic variables of the theory, those which
define the complete state description.  The latter may either in fact be
strongly primitive, or, like the electromagnetic field in classical
electrodynamics, they may be required in order to express the laws which
govern the behavior of the strongly primitive variables in a simple and
natural way. While this probably does not go far enough---we should further
distinguish those weakly primitive variables which, like the velocity, are
functions of the trajectory of the strongly primitive variables, and those,
again like the electromagnetic field, which are not---these details are not
relevant to our present purposes, so we shall ignore these distinctions.
\medskip

\noindent\note\ The change in experiment proposed by Albert is that 
``the {\it hardness box\/} is {\it flipped over\/}.'' However, with 
regard to spin this change will produce essentially no change in \(X\) 
at all.  To obtain the reversal of sign, either the polarity or the 
geometry of the SG magnet must be reversed, but not both.
\medskip

\noindent\note\ The classical references on this topic are: 
von~Neumann 1932, Gleason 1957, Jauch et al. 1963, Kochen et al. 1967.  
For a critical overview see Bell 1966, 1982.

\medskip

\noindent\note\ This is perhaps a bit too strong: As is well known, 
Bell (Bell 1964) has shown that no-hidden-variables-type arguments, suitably 
applied, can be used to establish the rather striking {\it physical\/} 
conclusion that nature is nonlocal.
\medskip

\noindent\note\ For some steps in the direction of the formulation of 
a Lorentz invariant Bohmian theory, as well as some reflections on the 
problem of Lorentz invariance, see Berndl et al. 1996.
\medskip

\noindent\note\  Everett 1957. See also De Witt et al. (eds.) 1973.
\medskip

\noindent\note\  We are referring here to Bell's reformulation of 
Everett's theory (Bell 1981).
\medskip

\noindent\note\ `Quantum logic' was proposed by Birkhoff and von Neumann 
(Birkhoff et al. 1936). For more recent presentations and developments,
see, e.g., Jauch 1968, and Beltrametti et al. 1981. 
\medskip

\noindent\note\ See Albert 1992.
\medskip

\noindent\note\ That quantum mechanics has to do with a sort of 
`noncommutative' probability originated probably with 
von Neumann 1932. A comprehensive list of the recent literature
 would probably be out of place here.
\medskip

\noindent\note\ We are referring to the so called GRW-theory (Ghirardi 
et al. 1986, 1990, 1995),  in particular, as presented by  Bell
(Bell 1987, p. 200). (See also the contribution of Ghirardi to this issue.)
\medskip

\noindent\note\ Kochen 1985, Dieks 1991, and van Fraassen 1991.
\medskip

\noindent\note\  Nelson 1966, 1985. (See also Goldstein 1987.)  
\medskip

\noindent\note\ Gell-Mann et al. 1993, Griffiths 1984, Omnes 1988.  

\section*{References}

\noindent Albert, D.Z.: 1992, {\it Quantum Mechanics and Experience\/}, 
Harvard University Press, Cambridge, MA.
\medskip

\noindent  Bell, J. S.: 1964, `On the Einstein-Podolski-Rosen 
paradox', {\it  Physics} {\bf 1}, 195--200. Reprinted in Bell 1987. 
\medskip

\noindent  Bell, J. S.: 1966, `On the problem of hidden variables in 
quantum mechanics', {\it Review of Modern Physics} {\bf 38}, 
447--452. Reprinted in Bell 1987. 
\medskip

\noindent  Bell, J. S.: 1981, `Quantum mechanics for cosmologists',
in {\it Quantum Gravity 2},  C. Isham, R. Penrose, and D.
Sciama (eds.),  Oxford University Press,  New York,
pp. 611--637.  Reprinted in Bell 1987.
\medskip

\noindent  Bell, J. S.: 1982, `On the impossible pilot wave', 
{\it Foundations of Physics} {\bf 12}, 989--999. Reprinted
in Bell,1987. 
\medskip

\noindent Bell, J. S.: 1987, {\it Speakable and unspeakable in
quantum mechanics\/}, Cambridge University Press,  Cambridge.
\medskip

\noindent Bell, J.  S.: 1990, `Against ``measurement'' ', {\em Physics 
World} {\bf 3}, 33--40.  [Also appears in `Sixty-two Years of 
Uncertainty: Historical, Philosophical, and Physical Inquiries into 
the Foundations of Quantum Mechanics', Plenum Press, New York, pp.  
17--31.
\medskip

\noindent Beltrametti, E.G.,  Cassinelli, G.: 1981, {\it The Logic of 
Quantum Mechanics\/}, Reading, Mass..
\medskip

\noindent Berndl, K., D\"{u}rr, D., Goldstein, S., Zangh\`{\i}, N.: 
1996, `EPR-Bell Nonlocality, Lorentz Invariance, and Bohmian Quantum 
Theory', quant-ph/9510027, preprint.  (To appear in {\it Physical 
Review A}, April 1996.)
\medskip

\noindent Berndl, K., D\"{u}rr, D., Goldstein, S., Peruzzi G., 
Zangh\`{\i}, N.: 1995, `On the Global Existence of Bohmian Mechanics', {\it 
Communications in Mathematical Physics} {\bf 173}, 647--673.  
\medskip

\noindent Birkhoff, G., von Neumann, J.: 1936, `The logic of Quantum 
Mechanics', {\it Ann. Math.} {\bf 37}. 
\medskip

\noindent Bohm, D.: 1952, `A Suggested Interpretation of the Quantum 
Theory in Terms of `Hidden' Variables, I and II,' {\it Physical 
Review} {\bf 85}, 166--193.  Reprinted in Wheeler and Zurek 1983, pp.  
369--396.
\medskip

\noindent Bohm, D., Hiley, B.J.:1993, {\it The Undivided Universe: 
An Ontological Interpretation of Quantum Theory\/}, Routledge \& Kegan 
Paul, London.
\medskip

\noindent Daumer, M.,  D\"urr, D., Goldstein, S.  and Zangh\`{\i}, N.: 
1996, `On the role of operators in quantum theory', in preparation.
\medskip

\noindent Davies, E.B.: 1976, {\it Quantum Theory of Open Systems\/}, 
Academic Press, London.

\noindent De Witt, B. S., Graham, N. (eds.): 1973, {\it The 
Many-Worlds interpretation of Quantum Mechanics\/}, Princeton, N.J..
\medskip

\noindent Dieks, D.: 1991, `On some alleged difficulties  in 
interpretation of quantum mechanics', {\it Synthese} {\bf 86},
77--86.
\medskip

\noindent D\"urr, D., Goldstein, S.  and Zangh\`{\i}, N.: 1992, 
`Quantum Equilibrium and the Origin of Absolute Uncertainty', {\it 
Journal of Statistical Physics} {\bf 67}, 843--907; ``Quantum 
Mechanics, Randomness, and Deterministic Reality,''{\it Physics 
Letters A} {\bf 172}, 6--12.
\medskip

\noindent D\"urr, D., Goldstein, S.  and Zangh\`{\i}, N.: 1996,
`Bohmian Mechanics as the Foundation of Quantum Mechanics',
in {\it Bohmian Mechanics and Quantum Theory: An Appraisal\/},
J. Cushing, A. Fine and S. Goldstein (eds.), Kluwer Academic Press.
\medskip

\noindent Everett, H.: 1957, `Relative state formulation of quantum 
mechanics', {\it Review of Modern Physics\/} {\bf 29}, 454---462.
Reprinted in De Witt et al. 1973, and Wheeler et al. 1983.
\medskip

\noindent Gell-Mann, M. and Hartle, J.B.: 1993, `Classical Equations for
Quantum Systems', {\it Physical Review D\/} {\bf 47}, 3345--3382.  \medskip

\noindent Ghirardi, G.C., Rimini, A. and Weber, T.: 1986, `Unified Dynamics for
Microscopic and Macroscopic Systems', {\it Physical Review D\/} {\bf 34},
470--491.
\medskip

\noindent Ghirardi, G.C., Pearle, P. and Rimini, A.: 1990, `Markov Processes in
Hilbert Space and Continuous Spontaneous Localization of Systems of
Identical Particles', {\it Physical Review A\/} {\bf 42}, 78--89.
\medskip

\noindent Ghirardi, G.C., Grassi, R. and Benatti, F.:1995, `Describing the
Macroscopic World: Closing the Circle within the Dynamical Reduction
Program', {\it Foundations of Physics\/} {\bf 23}, 341--364.
\medskip

\noindent Gleason, A. M.: 1957, `Measures on the closed subspaces of 
a Hilbert space', {\it Journal of Mathematics and Mechanics\/} {\bf 
6}, 885--893.
\medskip

\noindent Griffiths, R.B: 1984, `Consistent Histories and the Interpretation of
Quantum Mechanics' {\it Journal of Statistical Physics\/} {\bf 36\/},
219--272.
\medskip

\noindent Goldstein, S.: 1987, `Stochastic mechanics and quantum 
theory', {\it Journal of Statistical Physics\/} {\bf 47}, 645--667.
\medskip

\noindent Jauch, J.M., Piron, C.: 1963, `Can hidden variables be 
excluded in quantum mechanics?',{\it Helvetica Phisica Acta\/} {\bf 
36}, 827--837.
\medskip

\noindent Jauch, J.M.: 1968: {\it Foundations of Quantum Mechanics},
Addison-Wesley,  Reading, Mass..
\medskip

\noindent Holland, P.R.: 1993, {\it The Quantum Theory of Motion\/},
Cambridge University Press, Cambridge.
\medskip

\noindent Kochen, S., Specker, E. P.: 1967, `The problem of hidden 
variables in quantum mechanics', {\it Journal of Mathematics and 
Mechanics\/} {\bf 17}, 59--87.
\medskip

\noindent Kochen, S.: `A new interpretation of quantum mechanics',
in {\it Symposium on the Foundations of Modern Physics\/},
P. Lathi and P. Mittelstaedt (eds.), World Scientific, Singapore.
\medskip

\noindent Mott, N.F: 1929, {\it Proceedings of the Royal Society A\/} 
{\bf 124\/}, 440.
\medskip

\noindent Nelson, E.: 1966 `Derivation of the Schr\"odinger equation from
Newtonian mechanics' {\it Physical Review} {\bf 150}, 1079--1085.
\medskip

\noindent Nelson, E.: 1985 {\it Quantum Fluctuations}, 
Princeton University Press, Princeton.
\medskip

\noindent Omn\`es, R.: 1988, `Logical Reformulation of Quantum Mechanics I',
{\it Journal of Statistical Physics\/} {\bf 53\/}, 893--932.
\medskip

\noindent von Neumann, J.: 1932,  {\it Mathematische Grundlagen der 
Quantenmechanik\/} Springer Verlag, Berlin. English translation:
1955,  Princeton University Press, Princeton.
\medskip

\noindent Schr\"odinger, E.: 1935, `Die gegenw\"artige Situation in der
Quantenmechanik,' {\it Die Naturwissenschaften\/} {\bf 23\/}, 807--812,
824--828, 844-849. [Also appears in translation as ``The Present Situation
in Quantum Mechanics,'' in Wheeler and Zurek 1983, pp. 152--167.]
\medskip

\noindent van Fraassen, B.: 1991, {\it Quantum Mechanics, an Empiricist
View\/}, Oxford University Press, Oxford.
\medskip

\noindent Wheeler, J.A.,  Zurek, W.H.  (eds.): 1983, {\it Quantum 
Theory and Measurement\/}, Princeton University Press, Princeton.

} 
\end{document}